\documentclass[
12pt,
a4paper,
reprint,
onecolumn,
]{revtex4-1}

\usepackage[T1]{fontenc}
\usepackage[latin9]{inputenc}
\usepackage{graphicx}
\usepackage{setspace}
\usepackage{amsmath}
\usepackage{enumerate}
\usepackage{textcomp}
\usepackage{amssymb}
\usepackage{relsize}
\usepackage{appendix}
\usepackage{babel}
\usepackage{amsfonts}
\setcitestyle{square}

\usepackage[a4paper, margin=3.4925cm]{geometry}
\newcommand{\coleq}{:=}

\begin{document}

\title{Volatile Memory Motifs:\\
Minimal Spiking Neural Networks}

\author{Fabio Schittler Neves}
\affiliation{Chair for Network Dynamics, Center for Advancing Electronics Dresden (cfaed) and Institute for Theoretical Physics, TU Dresden, 01062 Dresden, Germany}
\affiliation{Institute for Theoretical Physics, TU Dresden, 01062 Dresden, Germany}
\email{fabio.neves@tu-dresden.de}

\author{Marc Timme}
\affiliation{Chair for Network Dynamics, Center for Advancing Electronics Dresden (cfaed) and Institute for Theoretical Physics, TU Dresden, 01062 Dresden, Germany}
\affiliation{Institute for Theoretical Physics, TU Dresden, 01062 Dresden, Germany}
\affiliation{Cluster of Excellence Physics of Life, TU Dresden, 01062 Dresden, Germany}
\affiliation{Lakeside Labs, 9020 Klagenfurt am W{\"o}rthersee, Austria}
\email{marc.timme@tu-dresden.de}

\begin{abstract}
How spiking neuronal networks encode memories in their different time and spatial scales constitute a fundamental topic in neuroscience and neuro-inspired engineering. Much attention has been paid to large networks and long-term memory, for example in models of associative memory. Smaller circuit motifs may play an important complementary role on shorter time scales, where broader network effects may be of less relevance. Yet, compact computational models of spiking neural networks that exhibit short-term volatile memory and actively hold information until their energy source is switched off, seem not fully understood. Here we propose that small spiking neural circuit motifs may act as volatile memory components. A minimal motif consists of only two interconnected neurons -- one self-connected excitatory neuron and one inhibitory neuron -- and realizes a single-bit volatile memory. An excitatory, delayed self-connection promotes a bistable circuit in which a self-sustained periodic orbit generating spike trains co-exists with the quiescent state of no neuron spiking. Transient external inputs may straightforwardly induce switching between those states. Moreover, the inhibitory neuron may act as an autonomous turn-off switch. It integrates incoming excitatory pulses until a threshold is reached after which the inhibitory neuron emits a spike that then inhibits further spikes in the excitatory neuron, terminating the memory. Our results show how external bits of information (excitatory signal), can be actively held in memory for a pre-defined amount of time. We show that such memory operations are robust against parameter variations and exemplify how sequences of multidimensional input signals may control the dynamics of a many-bits memory circuit in a desired way.
\end{abstract}

\maketitle

\section{Introduction}	
Memory plays a fundamental role in biological, bio-inspired and abstract artificial computing systems. While in standard computers memory is usually implemented as discrete components \cite{FORTIER20031}, which can be addressed by other components as needed, in self-organized dynamical systems, computation and memory are typically intertwined, with computation and memory access taking place concurrently \cite{hopfield1982neural,neves2021Bio}. In particular, neural networks are composed of many discrete computing units called neurons with memory being stored in the network's connectivity itself. The result of a computation is the observable emergent  collective dynamics.

Due to its parallel and distributed nature, memory studies on neural networks have traditionally focused on large-scale systems, which not only exhibit a variety of emerging phenomena but also may become analytically tractable in the limit of large networks ($N\rightarrow \infty$) \cite{Hertz1991Introduction}. One key example of such collective phenomena is associative memory, in which memories are represented as attractors in state space, such that partial initial information or corrupted input signals may be sufficient to recover original stored  associated memories. Most neural networks models, for memory or computation, are also non-volatile \cite{coolen2005Theory, burr2017neuromorphic}, such that the information stored in the connections need not to be actively maintained but stay long-term without ongoing energy inputs required. Volatility, however, may be essential for a variety of cognitive functions, as working memory and real-time planning \cite{Shiffrin1993Short,BARAK2014Working}, in particular on shorter timescales, where broader network effects may be of a lesser relevance.

\begin{figure}[b]
\begin{centering}
\includegraphics[width= 9cm,angle=0]{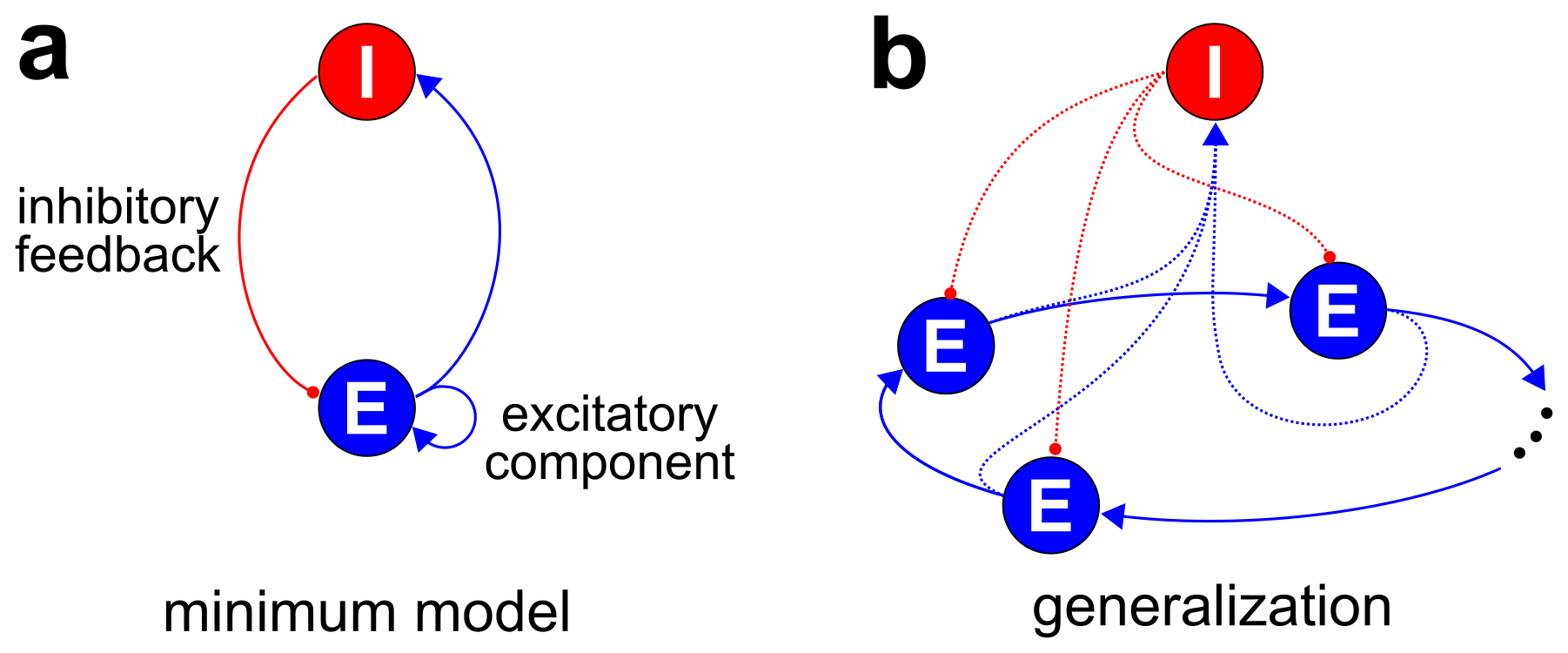}
\par\end{centering}
\caption[1st-entry]{\textbf{Spiking neural circuit motifs that implement a 1-bit volatile memory.} Blue circles labeled as `E' represent excitatory neurons, red circles labeled as `I' represent inhibitory neurons, circles as arrow heads represent inhibitory connections and conventional arrows represent excitatory connections. \textbf{(a)} A minimal circuit composed of two neurons, one excitatory and one inhibitory. \textbf{(b)} A circuit composed of an excitatory ring sub-network and an inhibitory neuron. \textbf{(a-b)} In both cases, the excitatory component has self-connections and connections to the inhibitory neurons while the inhibitory feedback connects to all excitatory neurons.}
\label{fig:system}
\end{figure}

Here we propose spiking neural network motifs to act as volatile memory components. A minimal example is composed of only two interconnected neurons (see Figure~\ref{fig:system}), one  excitatory neuron with a delayed self-connection (autapse) and one inhibitory neuron, yielding a bistable motif circuit. A self-sustained periodic spike-train, representing an `on' state and thus a bit `1' co-exists with the quiescent state, representing an `off'-state and thus a bit `0'. Switching between those states is controlled by transient external inputs to either the inhibitory or the excitatory neuron. Alternatively, the inhibitory neuron may also act as an autonomous off switch for the circuit. That neuron integrates the pulses incoming from the excitatory neuron until it reaches a spiking threshold upon which the inhibitory neuron emits a spike and terminates the self-sustained periodic spike-train at the excitatory neuron, overall turning the collective motif state from `1' to `0' . Collections of such motif may be used in parallel to represent more complex information as independent bits, if larger-scale network effects are not desirable or not relevant .

Our results below show how an external bit of information can be actively held in memory for a pre-defined amount of time. To hold a `1' bit in memory, neural spiking activity and thus energy is needed, making the memory system volatile. The small neural circuits introduced below may serve as basic memory units for short-term volatile memory, and thus may complement the broad variety of previously proposed computational neural circuits and memory models \cite{hebb2002Organization,coolen2005Theory,Langille2018Synaptic}, in particular the set of, also volatile, computing paradigms emerging in symmetrical systems \cite{Borresen2005Discrete, Wordsworth2008Spatiotemporal,Neves2012Computation,Neves2020Reconfigurable} or stochastic dynamics in random networks with local excitation and global inhibition \cite{Laing2001Stationary,Sandamirskaya2014dynamics}, which also take advantage of self-organization instead of carefully tuned (many) connections between neurons.  

\section{Minimal model of neuronal network motif}
\label{sec:model}
In this work we present a compact neuronal circuit motif that implements a 1-bit volatile memory. Volatile in this context means that spike activity, and thus energy, is needed to maintain at least one of the memory states. The system is compact because it can be implemented with as few as two neurons. For instance, as we sketch in  Figure~\ref{fig:system}a, the system may be implemented with one inhibitory neuron and one excitatory neuron only. The connectivity is such that the excitatory neuron has a self connection and a connection to the inhibitory neuron; the inhibitory neuron has a single connection to the excitatory neuron. Alternatively, the excitatory component may be composed of a ring (Figure~\ref{fig:system}b) or other small population of neurons to better resemble a biological neural circuit or to otherwise circumvent self-connections. To explain the basic mechanisms and collective dynamics underlying volatile memory function of such motifs, we consider a minimal motif of two neurons in the remainder of this article.

For clarity of presentation we here mathematically describe the neurons as Leaky Integrate-and-Fire neuron models that exhibit parameters with a direct physical meaning also for potential hardware implementations. Leaky integrate-and-fire models already capture main fundamental features of spiking neurons, including their dynamics exhibiting two different times scales: a  long term sub-threshold dynamics and short term interactions (spikes) modeled via discrete pulse responses. Our specific model is defined by a pair of differential equations,
\begin{eqnarray}
\label{eqn:exc}
\frac{dV_{E}}{dt} &=& A_{E} + \xi_{E}(t) - \gamma_{E} V_{E} + \sum_{t_{i} \in P_{E}} \varepsilon_{E} \delta(t-t_{i}-\tau_{E}) + \sum_{t_{j} \in P_{I}} \varepsilon_{I} \delta(t-t_{j}-\tau_{I})  + \eta_{E}(t) \\
\label{eqn:inh}
\frac{dV_{I}}{dt} &=& A_{I} + \xi_{I}(t) - \gamma_{I} V_{I} + \sum_{t_{i} \in P_{E}} \varepsilon_{E} \delta(t-t_{i}-\tau_{E}) + \eta_{I}(t) 
\end{eqnarray}
complemented by conditions for spike emission and reset. Specifically, we say that neuron $X\in\{E,I\}$ emits a spike at time $t:=t_n$ if its voltage reaches a threshold, $$V_{X}(t)\geq \theta$$
after which that voltage is reset to
$$V(t^+) \coleq 0.$$ 
The time $t_n$ indicates the $n$th spike time in the motif circuit (after some reference time $t_0$).
Moreover, the parameters $A_X$ represent temporally the internal driving currents that set the equilibrium voltage (see below), $\xi_{X}(t)$ external driving currents serving as input signal to store or remove memories, and $\gamma_{X}$ the leak constants. Finally $\varepsilon_{X}$ represent the connection weights, $\tau_{X}$ the delays between a spike emitted by neuron $X$ and reception of that spike and $P_{X}$ denotes the set of all pulses elicited by a neuron $X$. The indices $E$ and $I$ indicate features of the excitatory and inhibitory neurons, respectively. Finally the inputs $\eta_{X}$ are the contribution of internal noise. We remark that the autonomous part of the system of equations above, i.e. for $\eta_{X}(t)=\xi_{X}(t)\equiv 0$, has an analytical solution in between spike events and equally enable a piecewise, exact event-based simulations \cite{Mirollo1990Synchronization, hansel1998numerical, timme2006Speed}.

\section{Self-sustained and self-terminated memory}

\begin{figure}[b]
\begin{centering}
\includegraphics[width=12.5cm,angle=0]{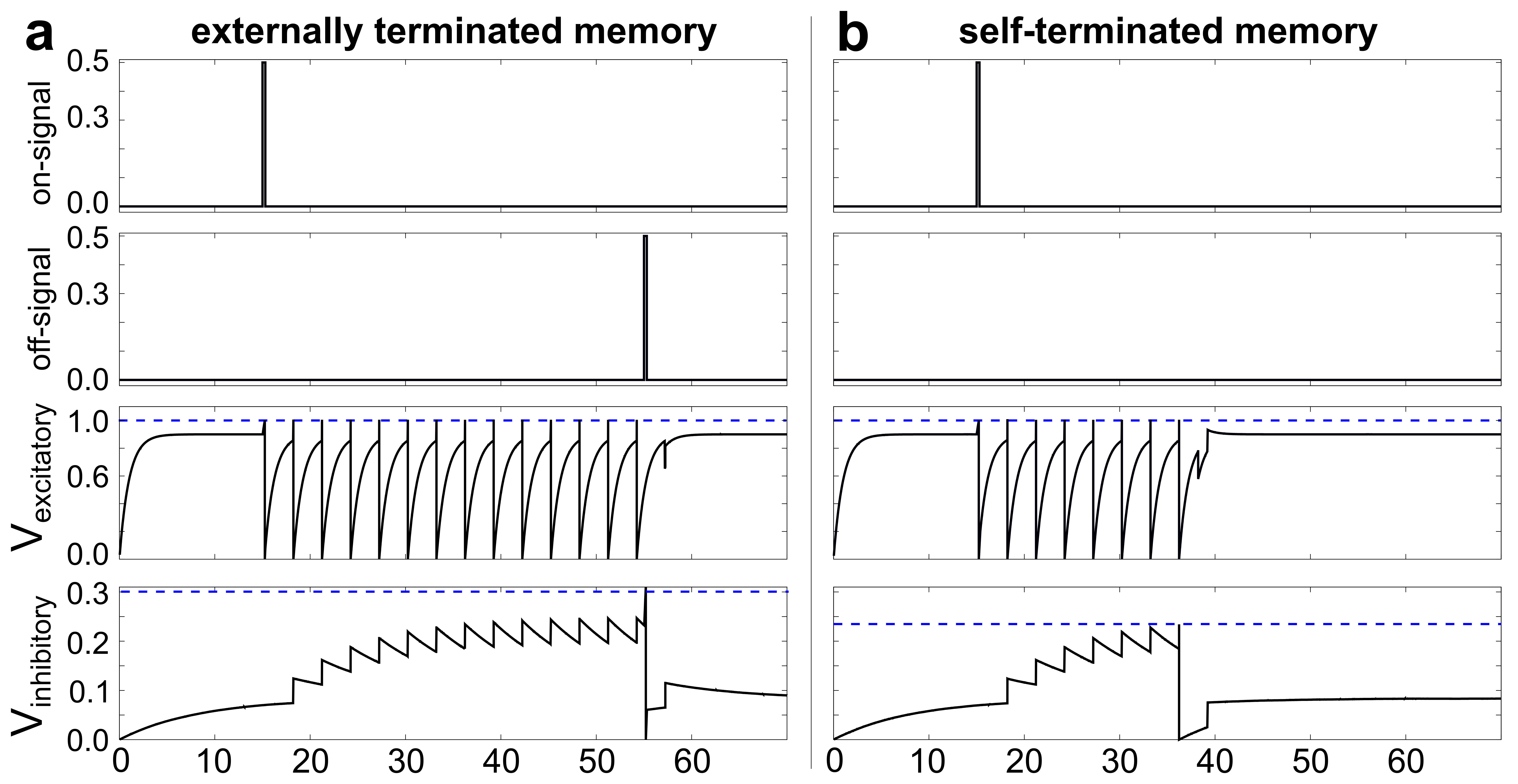}
\par\end{centering}
\caption[1st-entry]{\textbf{Bistable dynamics: memory initiation and termination.} For both panels, the upper graphs represent input currents as a function of time while the lower ones represent the voltages of inhibitory and excitatory neurons.\textbf{(a)} After a short input signal (upper panel), the excitatory neuron switches from its quiescent state to a self-sustained active state.A second external signal drives the inhibitory neuron to spike, which in turn terminates the memory, back to the quiescent state. \textbf{(b)} After the memory is initiated, the excitatory feedback loop persists until the inhibitory neuron produces a pulse, triggered by the consecutive excitatory pulses, thus terminating the memory. Parameters are: $A_E=0.9$, $A_I=0.01$, $\gamma_E=1$, $\gamma_I=0.12$, $\theta_E=1$, $\theta_I=0.3$, $\tau_E=3$, $\tau_I=2$, $\varepsilon_E=0.05$, $\varepsilon_I=-0.2$.}
\label{fig:dynamics}
\end{figure}

In the 2-unit motif network, two qualitatively different collective dynamics coexist (Figure~\ref{fig:dynamics}), one exhibiting a self-sustained spike train created by the excitatory neuron, encoding a bit value `1', the other a quiescent state with no spikes emitted, encoding the bit value `0'. To store the value `1', an excitatory signal $\xi_{E}$ is sent to the excitatory neuron; to switch from a bit `1' to zero, an external excitatory signal $\xi_{I}$ is sent to the inhibitory neuron. Both types of external signals  $\xi_{X}$ need to be sufficiently strong, i.e. of sufficiently large amplitude and duration. The exact shape of these input signals as a function of time is not relevant as long as they are sufficiently rapid and charge the targeted neuron sufficiently for it to cross threshold and spike.

In the absence of external input signals, the voltage of both neurons with time tends towards their respective fixed points $V_{E}\coleq I_{E}/\gamma_{E}$ and $V_{I}\coleq A_{I}/\gamma_{I}$,  see Figure~\ref{fig:dynamics} before input onset. A short transient input signal $\xi_{E}$ triggers the release of the first spike by the excitatory neuron. In turn, this spike arrives in both neurons after a delay $\tau_{E}$. For sufficiently strong pulse (response) amplitudes $\varepsilon_{E}$, the excitatory neuron sends a second spike and the process repeats. The motif network then maintains a spike-train with frequency $1/\tau_{E}$ until it is interrupted by an inhibitory pulse. There are two different mechanisms potentially causing such an interruption.
First, a strong excitatory signal $\xi_{I}(t)$ could be sent at any desired time from outside the motif, see Figure~\ref{fig:dynamics}a. Second, these systems hold the option of self-sustained  and self-terminating memory function  (see Figure~\ref{fig:dynamics}b), with memory duration set by system parameters (that might, in turn, be varied on demand): The ongoing sequence of excitatory pulses fed into the inhibitory neuron promotes consecutive voltage jumps. If one such spike brings the inhibitory neuron to or beyond its firing threshold, the inhibitory neuron elicits a spike that after a delay $\tau_{I}$ causes a voltage leak in the excitatory neuron, thereby  interrupting the  self-sustained spike-train. 

\section{Memory duration}	
An interesting feature of the memory circuit motif presented is its tunable memory duration. Quantitatively, how long the an on-state is held active before self-terminating depends on most of the system parameters, for example on the pulse amplitudes (and durations), the delays, and the leak constant $\gamma_{X}$. For a qualitative analysis, we study the memory duration in terms of variations of the leakage parameter $\gamma_{I}$ and the firing threshold $\theta_{I}$, fixing all other parameters. A natural way to measure the memory duration is in terms of number of elicited spikes; the absolute real time again depends on chosen parameters set in any motif implementation. Furthermore, because the excitatory neuron's role is simply to generate a spike-train with a fixed frequency $1/\tau_{E}$, we here studied the memory duration from the perspective of the inhibitory neuron's response to such spike-trains. 

\begin{figure}[b]
\begin{centering}
\includegraphics[width=9cm,angle=0]{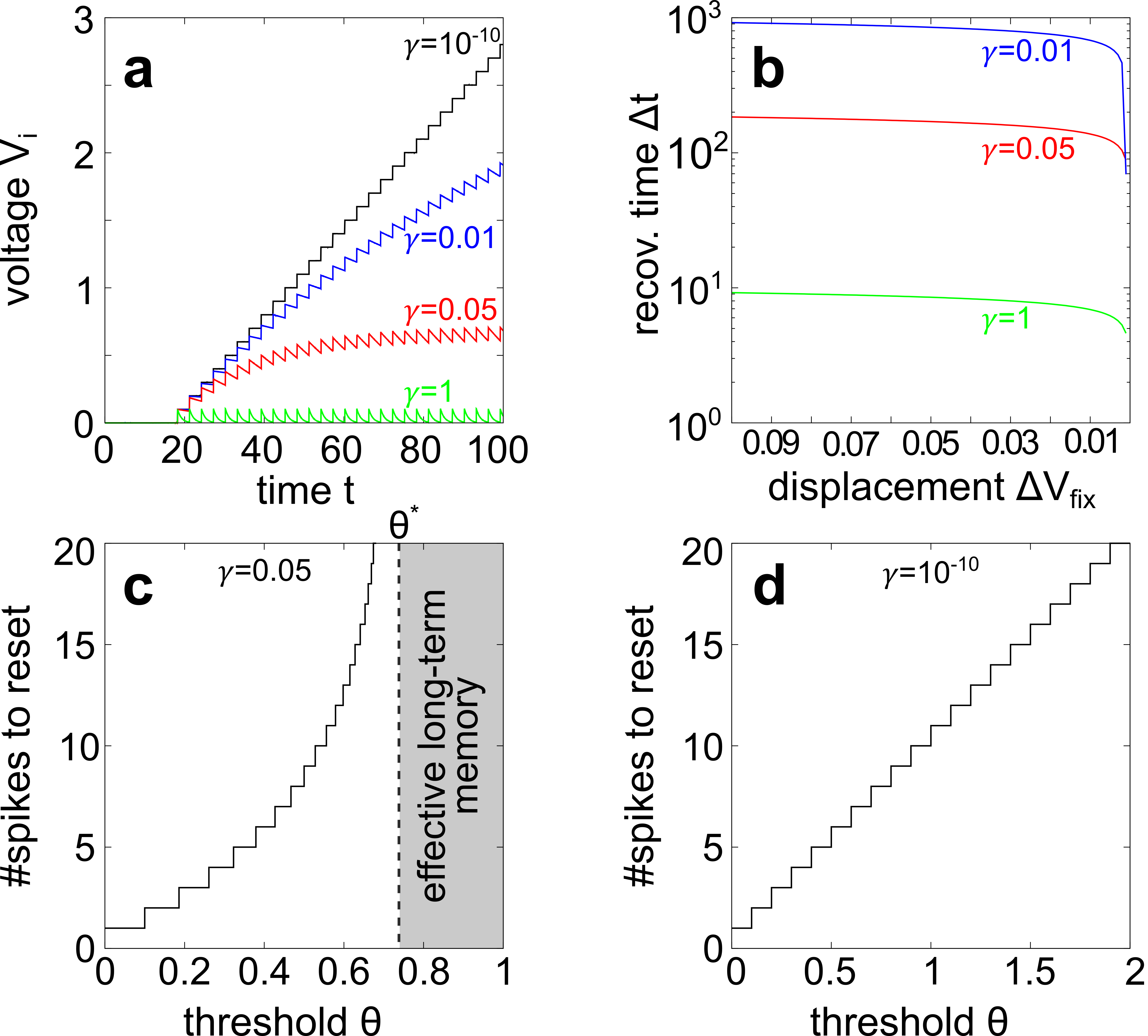}
\par\end{centering}
\caption[1st-entry]{\textbf{Memory duration and long-term dynamics} \textbf{(a)} Dynamical response of the inhibitory neuron to long spike trains with fixed frequency and amplitude. The larger the leak constant $\gamma_{I}$, the slower the voltage increases on average. \textbf{(b)} Given an instantaneous voltage jump $\Delta V_{fix}$ at time $t_0$ from the resting state, this panel shows the recovery time $\Delta t$, in which the voltage is $V(\Delta t)-V_{I} = 10^{-5}$ away from its resting states. The larger the leak constant $\gamma_{I}$, the shorter the recovery time for the same $\Delta V_{fix}$. \textbf{(c)} Number of spikes received by the inhibitory neuron until reset as a function of its threshold $\theta$. \textbf{(d)} As in \textbf{(c)}, but for a smaller $\gamma=10^{-10}$. The smaller $\gamma_{I}$ , the longer the curve resembles an equally spaced stair. Parameters: the same as in Figure~\ref{fig:dynamics} if not stated otherwise. $A_I=10^{-4}$ for all panels.}
\label{fig:memory}
\end{figure}

As shown in Figure~\ref{fig:memory}a, if $\gamma_{I}$ is large enough, most of the current injected into a neuron is lost during the inter-spike intervals and the voltage curve resembles a non-linear saw wave with a small up-drift. Contrariwise, in the limit of $\gamma_{I} \rightarrow 0$, no current is lost, as there is no leak term, and the voltage curve thus has a stair shape. Intermediary values show an average logarithmic increase overlayed by the spikes. Notice the exact values of $\gamma_{I}$ shown in Figure~\ref{fig:memory} are only illustrative, as the same qualitative effects can be achieve for fixed $\gamma_{I}$ and varying, for example, $\tau_{E}$ instead. We also expect the system to show robustness to small variations of $\varepsilon_{E}$ as the leakage grows exponentially with the deviation from the resting state, see Figure~\ref{fig:memory}b.

The memory duration is controllable. For depiction, we varied the firing threshold $\theta_{I}$ and fixed all other parameters. Figure~\ref{fig:memory}c shows how larger $\gamma_{I}$ values restrict the discernible memories' duration to a predefined interval of pulses, as the voltage peaks immediately after consecutive spikes get closer exponentially fast. For large enough $\theta_{I}$ the memory duration is in practice long-term, as the actual difference between the voltage peaks (at spike times) decreases exponentially with $\theta_{I}$ and the voltage does not converge to the threshold in finite time or converge to a value below the threshold. For $\gamma_{I} \rightarrow 0$, the leak is negligible, and the memory duration increases in equally spaced steps, multiples of to the spike amplitude $\varepsilon_{E}$ (Figure~\ref{fig:memory}d). We remark that long-term memory in this context does not imply storage but long-lasting, as the memory here is always sustained by pulses and is, thus, still volatile.

\section{Influence of Noise}	

\begin{figure}[b]
\begin{centering}
\includegraphics[width=11cm,angle=0]{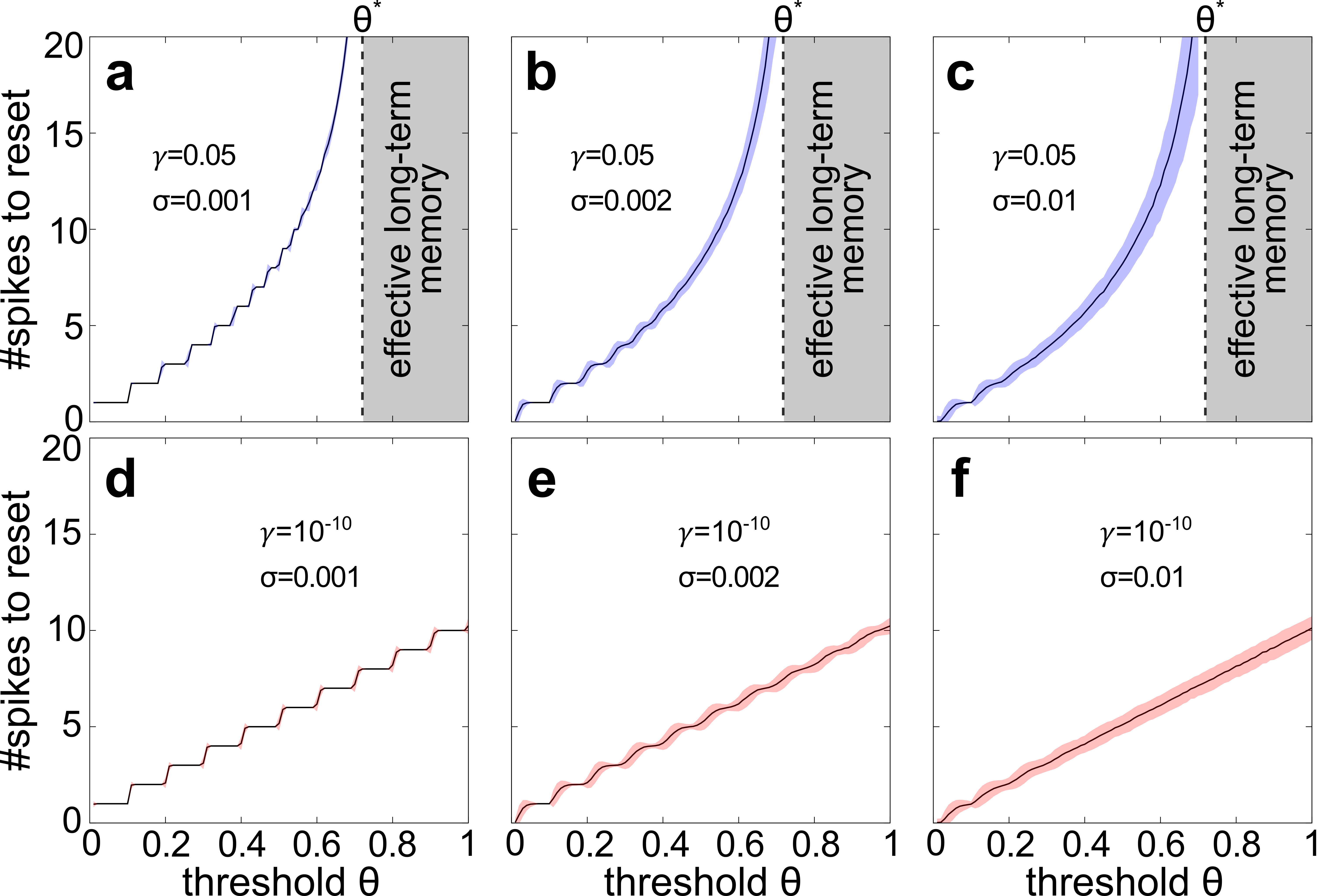}
\par\end{centering}
\caption[1st-entry]{\textbf{Noise-induced variability of memory duration.} Average number of spikes (black curves) received by the inhibitory neuron until its first spike and standard deviation (blue and red backgrounds) as functions of the inhibitory neuron's firing threshold. Measures calculated over $1000$ repetitions at intervals of $0.01$ volts. \textbf{(a-c)} For fixed $\gamma_{I}$, changing the noise standard deviation $\sigma$ softens the staircase features of the curve. Even though the mean does not seem to deviate too far from the noiseless case, the standard deviation monotonically increases  with $\theta$. \textbf{(d-e)} Same results as in \textbf{(a-c)}, but with a slower increase rate. Parameters: same parameters as in Figure~\ref{fig:memory} if not stated otherwise.}
\label{fig:noise}
\end{figure}

Noise is an ever-present feature in biological and artificial (hardware) neural networks.  Its role in memory and computation is varied and is often beneficial, in contrast to their effects in signal transmission lines. We now describe the effects of noise on our compact memory circuit. In the following, we add independent  Gaussian  noise sources  $\eta_{X}(t)$, with identically distributed random components, to both neurons. The noise is modeled (approximated) by adding the term
\begin{equation}
    \eta_{X}(t+\delta t)- \eta_{X}(t) =  \sqrt{\delta t} N_{rand}(\sigma,0)
\end{equation}
to the righthand side of equations (\ref{eqn:exc}) and (\ref{eqn:inh}). Here $N_{rand}(\sigma,0)$ is a random number drawn from a Gaussian distribution with variance $\sigma$ and centered at zero. To conserve the event-based feature time evolution
noise is evaluated after discrete times intervals $\delta t$ drawn independently from a Poisson distribution with average $\left<\delta t\right> = \tau_{E}/100$. That is, the noise sample intervals are randomized and independent for each neuron, while their average sampling interval is fixed.

Figure~\ref{fig:noise} illustrates the effect of different noise amplitudes for small and for intermediate noise amplitudes, $\gamma_{I} = 0.05$ and $\gamma_{I}=10^{-10}$, respectively. 
Qualitatively, the results are the same in their most important aspects. Initially, for small $\sigma$, noise mostly affects the vicinity of the transition points between two memories duration (step jumps). As a result, the steps themselves become less steep than without noise. For larger noise strength $\sigma$, the plateaus (progressively) lose their identity, due to the large variability in the memory duration between random realizations, as also reflected in an increase in the standard deviation of the number of spikes to reset. Because the voltage plateaus for large $\gamma_{I}$ become progressively smaller without noise, the increase in standard deviation becomes also progressively larger and more apparent with larger $\theta_{I}$, compare Figure~\ref{fig:noise}b-c to Figure~\ref{fig:noise}e-f. Furthermore, combinations of small enough $\theta_{I}$ with large enough $\sigma$ may promote eventual noise-induced spikes in the inhibitory neuron (false positive), even without an external signal to any of the neurons. As a consequence, spike events occur even before the excitatory neuron has its first spike, which translate in average memory duration below $1$ in Figures~\ref{fig:noise}(b-c) and Figures~\ref{fig:noise}(e-f). Notice, $1$ is the minimum memory length for the noise-free dynamics, that is, a single excitatory spike promoting an inhibitory spike.

\section{Loading and flushing multi-dimensional memories}
\begin{figure}[b]
\begin{centering}
\includegraphics[width=8cm,angle=0]{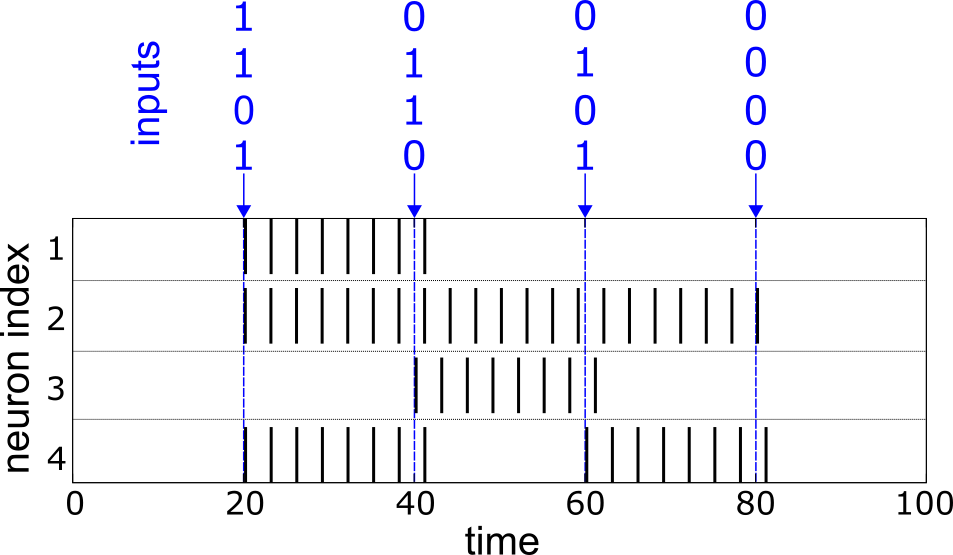}
\par\end{centering}
\caption[1st-entry]{\textbf{A sequence of four-bits words.} Only excitatory spikes are depicted. Four independent 1-bit neural circuit receive a sequence of four four-bits words. The last signal also serves to reset the system. Inputs label as $0$ represent short inputs to the inhibitory neuron and as $1$ to the excitatory neuron. In both cases the signal's duration is $0.3$ and the amplitude is $0.5$, see Figure~\ref{fig:dynamics} for details.}
\label{fig:bits}
\end{figure}

In neuronal systems, sets of neurons or neuronal networks can be used to represent and store information. In our approach to transiently holding bits in memory, 
neurons are interconnected forming small motifs. In the simplest setting, multi-dimensional memories may thus be established by multiple motifs acting independently and in parallel. In such settings, multi-dimensional inputs can be loaded concurrently into memory as independent bits, not unlike in a traditional computer (see Figure \ref{fig:bits}). In our model, loading a bit in memory is intuitive and in line with traditional computer, each single bit can be set to one of two states almost instantaneously (within one spike cycle interval), independently of the current states of the neuron. Furthermore, this system exhibits a natural ground state (non-active), to which the system abruptly switch after the memory interval elapses. Moreover, the state representation is very convenient for binary codes, as one state has spike activity and the other has a complete absence of spikes, thus, it does not require involved decoding approaches.

Figure \ref{fig:bits} shows how a sequence of words can be loaded into an array of neurons. As expected from our previous discussion around Figure \ref{fig:dynamics}, a new value can be loaded independent of the system state. The single false positive spikes after each active-to-quiescent change of states occurs due to the delay $\tau_{I}$, i.e. spike signals still in transit (sent but not yet received). As a consequence, the desired system state, i.e. the collective dynamics at all motifs, is assumed after a (small) lag time $\Delta t>\tau_{I}$.  The exact timing depends on the excitatory neuron's voltage at the time it receives the inhibition. This observation sets a minimum of two consecutive pulses with frequency $1/ \tau_{E}$ to guarantee a correct solution, because after a change in input signals, a single pulse may still be triggered by the former input signal (a false `1'). The real-world (clock) time interval $\tau_{E}$, e.g. in seconds, is defined by the neuron model time scales, by choosing the units for $\tau_{E}$.

\section{Discussion} 
We have proposed a general concept of implementing tunable volatile memory in simple neural networks. Such networks are small network motifs and exploit bi- or multistability to realize memory dynamically. Memory duration is either determined by system parameters that set the time scale of self-terminating a memory, or by external signals.

The concept of volatile memory is already familiar in computer science, see \cite{Mishra2016Generic}. It is defined as a memory type that is actively maintained by the system, thus continuously consuming power. Contrary to storage, such memory is erased each time power is no longer provided to the system. Inspired by such concepts, we here proposed that simple neural motifs may act as volatile memory components. Our model is fundamentally different from previous neuronal models with similar functionality, which rely on, e.g., short-term plasticity (\cite{tetzlaff2013synaptic}), because our model requires no changes in the network connectivity or their weights. Instead, memory is held dynamically in the spike configuration until terminated internally or externally. We specifically analyzed a simple 1-bit volatile memory neural network motif that exhibit bi-stability. The bit `1' is represented by a self-sustained spike train and the bit `0' by no spiking activity. 

Our focus on minimal motifs was motivated by two aspects: first, independent bits may play an important role in small systems, where network effects may be less relevant; second, the minimal 2-neuron systems offers maximal clarity in gaining insights about fundamental mechanisms that underlie both the self-organized collective dynamics of a motif and its response to external control signals. We remark that the same concept and mechanisms underlie also volatile memory dynamics in larger recurrent motifs that exhibit a suitable inhibitory component (shutdown-counter) and may thus self-terminate memory. 
In general, for larger motifs or several motifs embedded into a larger network, future work will need to investigate two aspects, local memory function and broader network effects. Larger motifs or networks may also hold the option for additional, potentially more advanced, functionality, for instance into the direction of systematically correlated multi-bit parallel memory storage, see also \cite{maass2007computational, Neves2012Computation}.

We chose a standard leaky integrate-and-fire neuronal model \cite{stein1965theoretical, knight1972dynamics, teeter2018generalized} to keep the number of defining parameters to the most essential ones. Nevertheless, the conditions to implement such volatile memory circuit do not depend on the details of the neuronal model, but only on whether a self-sustained spike-train can be initiated by an external signal and whether the inhibitory feedback can promptly terminate such a spike-train. The results might thus be viewed as conceptual and largely independent of the neuron model.

Departing in some measure from the biological paradigm, independent bits (motif states) can be assembled to form larger sets of $N$ motifs which combined have a large memory capacity ($2^N$), as in traditional computers. While it is unclear if the animal brain may take advantage of such combinatorial approach, bio-inspired computers can certainly make use of it to complement functionality of a large class of spiking neural systems, thereby maintaining information and processing completely within the spiking paradigm if desired.  

Our minimal motif for volatile memory complements a variety of alternative dynamical system models of neural and networked information processing systems \cite{Ashwin2005When,rabinovich2001dynamical,rabinovich2006dynamical,lukovsevivcius2009reservoir}. In particular, our model for short-term memory is a promising complement for approaches to computations relying on simple (neural) logical gates or on symmetrical spiking neural systems \cite{Borresen2005Discrete, Wordsworth2008Spatiotemporal,Neves2012Computation,Neves2020Reconfigurable}. To date, these systems transiently process information but cannot retain the result of a computation, neither in the long- nor in the short-term, for example in (noisy) heteroclinic networks \cite{Wordsworth2008Spatiotemporal} or, more generally, networks of unstable states \cite{Neves2012Computation}. Finally, we believe that such alternative and compact form of volatile memory implementation may contribute to future computing architectures, e.g., in neuromorphic and bio-inspired chemical, physical and robotic systems \cite{okeefe2019reviewSwarmalators, schilcher2021swarmalators, smelov2019controllable, Amirhossein2022Advancements}. 

\section*{Acknowledgements}
Partially supported by the German Research Foundation (Deutsche Forschungsgemeinschaft, DFG) under project number 419424741 and under Germany's Excellence Strategy -- EXC-2068 -- 390729961 -- Cluster of Excellence Physics of Life at TU Dresden, and the Saxonian State Ministry for Science, Culture and Tourism under grant number 100400118.

\bibliographystyle{unsrt}
\bibliography{bibliography}

\begin{thebibliography}{10}

\bibitem{FORTIER20031}
P~J Fortier and H~E Michel.
\newblock 1 - introduction.
\newblock In P~J Fortier and H~E Michel, editors, {\em Computer Systems
  Performance Evaluation and Prediction}, pages 1--38. Digital Press,
  Burlington, 2003.

\bibitem{hopfield1982neural}
J~J Hopfield.
\newblock Neural networks and physical systems with emergent collective
  computational abilities.
\newblock {\em Proceedings of the National Academy of Sciences},
  79(8):2554--2558, 1982.

\bibitem{neves2021Bio}
F~S Neves and M~Timme.
\newblock Bio-inspired computing by nonlinear network dynamics{\textemdash}a
  brief introduction.
\newblock {\em Journal of Physics: Complexity}, 2(4):045019, 2021.

\bibitem{Hertz1991Introduction}
J~Hertz, A~Krogh, and R~G Palmer.
\newblock {\em Introduction to the Theory of Neural Computation}.
\newblock Westview Press, 1991.

\bibitem{coolen2005Theory}
A~C~C Coolen, R~K{\"u}hn, and P~Sollich.
\newblock {\em Theory of Neural Information Processing Systems}.
\newblock Oxford University Press, 2005.

\bibitem{burr2017neuromorphic}
G~W Burr, R~M Shelby, A~Sebastian, S~Kim, S~Kim, S~Sidler, K~Virwani, M~Ishii,
  P~Narayanan, A~Fumarola, et~al.
\newblock Neuromorphic computing using non-volatile memory.
\newblock {\em Advances in Physics: X}, 2(1):89--124, 2017.

\bibitem{Shiffrin1993Short}
R~M Shiffrin.
\newblock Short-term memory: A brief commentary.
\newblock {\em Mem. Cogn.}, 21:193--197, 1993.

\bibitem{BARAK2014Working}
O~Barak and M~Tsodyks.
\newblock Working models of working memory.
\newblock {\em Current Opinion in Neurobiology}, 25:20--24, 2014.

\bibitem{hebb2002Organization}
D~O Hebb.
\newblock {\em The Organization of Behavior: A Neuropsychological Theory (1st
  ed.)}.
\newblock Psychology Press, 2002.

\bibitem{Langille2018Synaptic}
J~J Langille and R~E Brown.
\newblock The synaptic theory of memory: A historical survey and reconciliation
  of recent opposition.
\newblock {\em Frontiers in Systems Neuroscience}, 12, 2018.

\bibitem{Borresen2005Discrete}
P~Ashwin and J~Borresen.
\newblock Discrete computation using a perturbed heteroclinic network.
\newblock {\em Phys. Lett. A}, 374(4--6):208--214, 2005.

\bibitem{Wordsworth2008Spatiotemporal}
J~Wordsworth and P~Ashwin.
\newblock Spatiotemporal coding of inputs for a system of globally coupled
  phase oscillators.
\newblock {\em Phys. Rev. E}, 78:066203, 2008.

\bibitem{Neves2012Computation}
F~S Neves and M~Timme.
\newblock Computation by switching in complex networks of states.
\newblock {\em Phys. Rev. Lett.}, 109(1):018701, 2012.

\bibitem{Neves2020Reconfigurable}
F~S Neves and M~Timme.
\newblock Reconfigurable computation in spiking neural networks.
\newblock {\em IEEE Access}, 8:179648--179655, 2020.

\bibitem{Laing2001Stationary}
C~R Laing and C~C Chow.
\newblock Stationary bumps in networks of spiking neurons.
\newblock {\em Neural Comput.}, 13(7):1473--94, 2001.

\bibitem{Sandamirskaya2014dynamics}
Y~Sandamirskaya.
\newblock Dynamic neural fields as a step toward cognitive neuromorphic
  architectures.
\newblock {\em Front. Neurosci.}, 7, 2014.

\bibitem{Mirollo1990Synchronization}
R~E Mirollo and S~H Strogatz.
\newblock Synchronization of pulse-coupled biological oscillators.
\newblock {\em SIAP}, 50(6):1645--1662, 1990.

\bibitem{hansel1998numerical}
D~Hansel, G~Mato, C~Meunier, and L~Neltner.
\newblock On numerical simulations of integrate-and-fire neural networks.
\newblock {\em Neural Computation}, 10(2):467--483, 1998.

\bibitem{timme2006Speed}
M~Timme, T~Geisel, and F~Wolf.
\newblock Speed of synchronization in complex networks of neural oscillators:
  Analytic results based on random matrix theory.
\newblock {\em Chaos}, 16:015108, 2006.

\bibitem{Mishra2016Generic}
S~Mishra, N~K Singh, and V~Rousseau.
\newblock Chapter 3 - generic soc architecture components.
\newblock In S~Mishra, N~K Singh, and V~Rousseau, editors, {\em System on Chip
  Interfaces for Low Power Design}, pages 29--51. Morgan Kaufmann, 2016.

\bibitem{tetzlaff2013synaptic}
C~Tetzlaff, C~Kolodziejski, M~Timme, M~Tsodyks, and F~W{\"o}rg{\"o}tter.
\newblock Synaptic scaling enables dynamically distinct short-and long-term
  memory formation.
\newblock {\em PLoS Computational Biology}, 9:e1003307, 2013.

\bibitem{maass2007computational}
W~Maass, P~Joshi, and E~D Sontag.
\newblock Computational aspects of feedback in neural circuits.
\newblock {\em PLoS Comput Biol}, 3(1):e165, 2007.

\bibitem{stein1965theoretical}
R~B Stein.
\newblock A theoretical analysis of neuronal variability.
\newblock {\em Biophysical Journal}, 5(2):173--194, 1965.

\bibitem{knight1972dynamics}
B~W Knight.
\newblock Dynamics of encoding in a population of neurons.
\newblock {\em The Journal of general physiology}, 59(6):734--766, 1972.

\bibitem{teeter2018generalized}
C~Teeter, R~Iyer, V~Menon, N~Gouwens, D~Feng, J~Berg, A~Szafer, N~Cain, H~Zeng,
  M~Hawrylycz, et~al.
\newblock Generalized leaky integrate-and-fire models classify multiple neuron
  types.
\newblock {\em Nature Communications}, 9(1):709, 2018.

\bibitem{Ashwin2005When}
P~Ashwin and M~Timme.
\newblock When instability makes sense.
\newblock {\em Nature}, 436:36--37, 2005.

\bibitem{rabinovich2001dynamical}
M~Rabinovich, A~Volkovskii, P~Lecanda, R~Huerta, H~D~I Abarbanel, and
  G~Laurent.
\newblock Dynamical encoding by networks of competing neuron groups: winnerless
  competition.
\newblock {\em Phys. Rev. Lett.}, 87(6):068102, 2001.

\bibitem{rabinovich2006dynamical}
M~I Rabinovich, P~Varona, A~I Selverston, and H~D~I Abarbanel.
\newblock Dynamical principles in neuroscience.
\newblock {\em Reviews of Modern Physics}, 78(4):1213, 2006.

\bibitem{lukovsevivcius2009reservoir}
M~Luko{\v{s}}evi{\v{c}}ius and H~Jaeger.
\newblock Reservoir computing approaches to recurrent neural network training.
\newblock {\em Computer Science Review}, 3(3):127--149, 2009.

\bibitem{okeefe2019reviewSwarmalators}
K~O'Keeffe and C~Bettstetter.
\newblock A review of swarmalators and their potential in bio-inspired
  computing.
\newblock {\em Micro-and Nanotechnology Sensors, Systems, and Applications XI},
  10982:383--394, 2019.

\bibitem{schilcher2021swarmalators}
U~Schilcher, J~F Schmidt, A~Vogell, and C~Bettstetter.
\newblock Swarmalators with stochastic coupling and memory.
\newblock In {\em 2021 IEEE International Conference on Autonomic Computing and
  Self-Organizing Systems (ACSOS)}, pages 90--99. IEEE, 2021.

\bibitem{smelov2019controllable}
P~S Smelov, I~S Proskurkin, and V~K Vanag.
\newblock Controllable switching between stable modes in a small network of
  pulse-coupled chemical oscillators.
\newblock {\em Physical Chemistry Chemical Physics}, 21(6):3033--3043, 2019.

\bibitem{Amirhossein2022Advancements}
A~Javanshir, T~T Nguyen, M~A~P Mahmud, and A~Z Kouzani.
\newblock {Advancements in Algorithms and Neuromorphic Hardware for Spiking
  Neural Networks}.
\newblock {\em Neural Computation}, 34(6):1289--1328, 2022.

\end{thebibliography}

\end{document}